\documentclass[openacc]{rstransa}%%%%where rstrans is the template name

\usepackage{bm}
\usepackage{subcaption} % Paquete recomendado para subfiguras
\usepackage{comment}

\newcommand{\R}{\mathbb{R}}

\newcommand{\C}{\mathbb{C}}
\newcommand{\Z}{\mathbb{Z}}

\newcommand{\D}{\Delta}
\newcommand{\vk}{\textbf{k}}
\newcommand{\vx}{\textbf{x}}
\newcommand{\vu}{\textbf{u}}
\newcommand{\vw}{\bm{\omega}}
\newcommand{\La}{\raisebox{1.5ex}{\scalebox{1}[-1]{$\mathbb{V}$}}} 
\newcommand{\re}{\text{Re}}
\newcommand{\vq}{\textbf{q}}
\newcommand{\proj}{\mathbb{P}}

\titlehead{Research}

\begin{document}

%%%% Article title to be placed here
\title{Spontaneous stochasticity in the fluctuating Navier-Stokes equations on a logarithmic lattice}

\author{%%%% Author details
Erika Ortiz$^{1}$, Ciro S. Campolina$^{2}$ and  Alexei A. Mailybaev$^{1}$}

%%%%%%%%% Insert author address here
\address{$^{1}$Instituto de Matemática Pura e Aplicada - IMPA, Rio de Janeiro, Brazil\\
$^{2}$Scuola Normale Superiore, Pisa, Italy}

%%%% Subject entries to be placed here %%%%
\subject{xxxxx, xxxxx, xxxx}

%%%% Keyword entries to be placed here %%%%
\keywords{Spontaneous stochasticity, fluctuating hydrodynamics, logarithmic lattice}

%%%% Insert corresponding author and its email address}
\corres{Erika Paola Ortiz Bernal\\
\email{erika.ortiz@impa.br}}

%%%% Abstract text to be placed here %%%%%%%%%%%%
\begin{abstract}
	The predictability of turbulent flows remains a challenging problem for mathematicians, physicists, and meteorologists.
	In this context, we consider the 3D incompressible Navier-Stokes equations with small-scale random forcing on logarithmic lattices in Fourier space.
	Our goal is to probe the phenomenon of spontaneous stochasticity in this system, which means that its solutions remain stochastic in the limit of vanishing viscosity and noise.
	For this, we consider numerical simulations with increasing Reynolds numbers and vanishing noise amplitudes.
	Through measurements of statistics of individual large-scale Fourier modes, we verify the spontaneous stochasticity in two different setups: from rough initial data, and after a finite-time blowup of a strong solution.
	The convergence of probability density functions for distinct parameters suggests that the limiting solution is a universal stochastic process.
\end{abstract}
%%%%%%%%%%%%%%%%%%%%%%%%%%%

%%%%%%%%%% Insert the texts which can accomdate on firstpage in the tag "fmtext" %%%%%

\begin{fmtext}
\end{fmtext}

%%%%%%%%%%%%%%% End of first page %%%%%%%%%%%%%%%%%%%%%

\maketitle

\section{Introduction}

The multi-scale and nonlinear nature of turbulent flows manifest in many physical phenomena~\cite{frisch1995turbulence}.
One of the most challenging problems of turbulence is to determine whether a turbulent flow remains predictable within a certain time horizon.
For a given finite Reynolds number, the flow is chaotic~\cite{deissler1986navier}, and the rate of growth of infinitesimal perturbations is dominated by the fastest time scale of the flow~\cite{ruelle1979microscopic}.
This means that the largest Lyapunov exponent is proportional to the inverse of Kolmogorov's time scale, and hence, it diverges with increasing Reynolds.
In his pioneering works~\cite{Lorenz1,Lorenz2}, Lorenz observed that such perturbations occurring at small scales could be progressively amplified to large integral scales within their characteristic turn-over times, in a backward cascade of error~\cite{leith1972predictability,eyink1996turbulence,boffetta1997predictability,boffetta2017chaos}.
Most strikingly, Lorenz noticed that some deterministic systems can evolve in such an unpredictable manner that they become practically indistinguishable from indeterministic (random) systems~\cite{Lorenz2}.
To illustrate this behavior, he analyzed pairs of solutions of a nonlinear fluid system which initially differed by a small observational error.
Such solutions evolved over time to completely different states, and this difference could not be reduced simply by decreasing the initial error, indicating that perfect initial accuracy was not sufficient to improve the system's predictability.
This phenomenon was originally called the butterfly effect, although later this name came to be used in the context of deterministic chaos~\cite{palmer2014real}.

In the last decades, this phenomenon received increasing attention, and has been reformulated under the name of "spontaneous stochasticity".
Among many competing definitions, the core idea is that the evolution of non-smooth (singular) solutions governed by deterministic equations remain intrinsically random in the double limit of vanishing regularization and vanishing noise.
The phenomenon has been studied in shell models of turbulence~\cite{SptS1,SptS2}, formalized for ordinary differential equations with non-Lipshitz singularities~\cite{eyink2020renormalization,drivas2021life,drivas2024statistical}, and numerically observed for the 2D Kelvin-Helmholtz shear instability~\cite{Butterfly}, this latter in a rather universal way.
Such universality found explanations using the renormalization group formalism~\cite{mailybaev2023spontaneous,mailybaev2023spontaneously,mailybaev2025rg}.

In all this context, highly turbulent flows under any perturbation---no matter how small---might loose predictability at all scales within a few integral turn-over times.
As a consequence, such reality would still be true when considering the unavoidable random perturbation from molecular noise.
Trying to explore this scenario, recent works~\cite{Bandak,Diss-Rate,Eyink,Eyink2} relied on the theory of fluctuating hydrodynamics, which incorporates the random microscopic effects of individual molecules to the continuum macroscopic description of the Navier-Stokes equations~\cite{Landau}.
The idea was to study how the effects of thermal noise could impact the large scale predictability in the Navier-Stokes equations.
Indeed, numerical simulations of a shell model with stochastic forcing confirmed the expected spontaneous stochastic behavior, in a parameter range compatible with the atmospheric boundary layer~\cite{Bandak}.

Despite all the above advances, spontaneous stochasticity has never been probed---not even numerically---for the 3D Navier-Stokes equations.
This is certainly due to the great computational complexity of the task, combining the incredibly large number of required degrees of freedom for each simulation, and the necessity of multiple (possibly thousands) of realizations.

Recently, the 3D Navier-Stokes system was formulated on logarithmic lattices~\cite{LogL}, \textit{i.e.} grid points increasing geometrically in Fourier space.
This technique retains many of the properties of the original system (like the exact form of the equations, the group of symmetries, the inviscid invariants, \textit{etc}), and still drastically reduces the number of degrees of freedom.
For this reason, numerical simulations of high Reynolds numbers are accessible and allow an accurate statistical study of solutions.
Our aim in the present work is to introduce the 3D Landau-Lifshitz-Navier-Stokes equations for fluctuating hydrodynamics on logarithmic lattices and study its infinite Reynolds limit, which corresponds to vanishing regularization and vanishing noise. 
We focus on the convergence of the solutions at large scales, at which we expect the convergence to be faster. Therefore, we measure the probability distribution functions of individual large scale modes and probe whether the limiting solutions remain stochastic or not.
Our results have physical relevance in the framework of homogeneous isotropic turbulence by describing the regime of large Reynolds numbers. Of course, this relevance is only qualitative, as the log-lattice models have considerably lesser number of degrees of freedom than the original Navier-Stokes system. However, observation of the spontaneous stochasticity in such models provides valuable insights that can later be extended to real-world turbulent flows.

The paper is organized as follows.
In Section~\ref{SEC:FluctuatingHydrodynamics}, we introduce the Landau-Lifshitz-Navier-Stokes equations for fluctuating hydrodynamics and formulate the spontaneous stochasticity hypotheses in this framework.
In Section~\ref{SEC:LogLatt}, we develop the logarithmic lattice model and explain the numerical methods.
In Section~\ref{SEC:results} we present the numerical results.
In Section~\ref{SEC:conclusions}, we address the conclusions.

\section{Fluctuating hydrodynamics}
\label{SEC:FluctuatingHydrodynamics}

The Navier-Stokes equations offer a mathematical description for the evolution of velocity and pressure fields of incompressible viscous flows.
However, this classical description abstracts the motion of individual molecules, which experience complex random collisions among each other, resulting in stochastic dynamics at microscales.
These two descriptions overlap at the mesoscale, since it is possible to describe a fluid using macroscopic field variables, but incorporating the effects of thermal fluctuations.
This motivates the extension of classical hydrodynamics through stochastic formulations.
This theory was first developed by Landau and Lifshitz~\cite{Landau}, and incorporates fluctuating terms into the Navier-Stokes equations in a way consistent with Statistical Mechanics.

Following this approach, we represent the velocity field $\vu(\vx,t)\in \R^3$ and the scalar pressure $p(\vx,t)\in \R$  with $\vx\in\R^3$ and $t\in\R$.
Then, the Landau-Lifshitz-Navier-Stokes equations for fluctuating hydrodynamics are~\cite{Landau,Diss-Rate}
\begin{equation}\label{NS_withNoise}
    \partial_t \vu + \vu \cdot \nabla\vu = -\nabla p + \frac{1}{Re}\Delta \vu + \nabla \cdot \bm{\tau}, \qquad \nabla\cdot \vu=0, 
\end{equation}
where $\bm{\tau}(\bm{x},t)= (\tau_{ij}) \in \R^{3\times 3}$ are the fluctuating stresses, modeled as a Gaussian random tensor, both symmetric and traceless, with zero mean and covariance given by
\begin{equation}\label{covarianza} 
    \langle \bm{\tau}_{ij}(\vx,t)\bm{\tau}_{kl}(\vx',t') \rangle = \Theta \left(\delta_{ik}\delta_{jl}+\delta_{il}\delta_{jk}-\frac{2}{3}\delta_{ij}\delta_{kl} \right) \delta^3(\vx-\vx')\delta(t-t').
\end{equation}
The dimensionless parameters are the Reynolds number $Re$ and the amplitude of noise $\Theta$, expressed as
\begin{equation}\label{Re_Theta}
	Re = \frac{UL}{\nu}, \qquad \Theta = \frac{2\nu k_B T}{\rho L^4 U^3},
\end{equation}
where $U$ is  the large-scale velocity of the flow, $L$ is the characteristic integral length, $\nu$ is the kinematic viscosity, $k_B$ is Boltzmann’s constant, $T$ is the absolute temperature, and $\rho$ is the mass density.

Covariance form~\eqref{covarianza} can be derived by demanding the fluctuating stresses $\bm{\tau}$ to be uncorrelated in space and time, to have isotropic statistics, and to keep the conservation of momentum and incompressibility of the model.
The constant $\Theta$ in Eq.~\eqref{Re_Theta} is obtained from fluctuation-dissipation theorems~\cite{DeZarate}.
We refer the reader to \ref{AppA} and references therein for the deduction of the model.

Giving precise mathematical sense of Eq.~\eqref{NS_withNoise} as a stochastic partial differential equation is however difficult, due to the additive noise growing in the small scales.
We can overcome this issue by recalling that the model has no physical validity at microscales, \textit{i.e.} at scales below the continuum hypothesis.
Mathematically, a high-wavenumber cutoff $\Lambda$ is introduced in Fourier space, and the system reduces to a classical Langevin-type equation for the Fourier modes~\cite{Eyink}.
The respective dimensionless parameter is defined as $\delta = 1/L\Lambda$, which represents the cutoff scale in units of the integral length.
The choice of the cut-off $\Lambda$ is somewhat arbitrary, but subjected to important constraints, since it should be in the range $\eta^{-1} \ll \Lambda \ll \lambda_{\text{mfp}}^{-1}$, where $\eta = \varepsilon^{-1/4}\nu^{3/4}$ is Kolmogorov dissipation scale and $\lambda_{\text{mfp}}$ is the mean free path of molecules.
Anyhow, the dimensionless cutoff scale $\delta =1/L\Lambda \to 0$ vanishes as $Re \to \infty$, because $\eta/L \propto Re^{-3/4}$ and $\lambda_{\text{mfp}}/L \propto Re^{-1}$.

 Using flow parameters characteristic of the atmospheric boundary layer (ABL)~\cite{ABL}, $T=300^\circ \,K$, $\nu=1.5\times 10^{-5}\, m/s$, $\rho=1.2 \,Kg/m^3$, $\varepsilon=4\times 10^{-2}\, m^2/s^3$, $L=10^3 \,m$ and $U=3.42 \, m/s$, we have $\Theta \simeq 10^{-39}$, $Re^{-1}\simeq 10^{-8}$ and $\delta \simeq 10^{-10}$.
 Such typical small values naturally suggest to take the triple limit $Re^{-1}, \Theta, \delta \to 0$ and justify the conventional perception that relevant turbulent fluid flows are well modeled by the deterministic incompressible Euler equations
 \begin{equation}
 	\partial_t \vu + \vu\cdot \nabla \vu = -\nabla p, \quad \nabla \cdot \vu = 0.
 \end{equation}

Because this is a triple limit, there is no unique way to arrive at it, and the way these limits are taken will influence the result.
If the noise is removed too quickly compared to the viscosity, the final limit may turn out to be deterministic, and so, spontaneous stochasticity would not be observed.
Furthermore, if appropriate scales are not considered, one may discard scales where relevant turbulent fluctuations may still exist or include microscopic scales where hydrodynamics can no longer be considered continuous~\cite{Diss-Rate}.
In order to deal properly with this triple limit, and still with certain generality, we take all parameters scaling with $Re$ as
\begin{equation}\label{triple_limit}
    Re^{-1}\to 0, \hspace{0.5cm} \Theta=\theta Re^{-\alpha}\to 0, \hspace{0.5cm}\text{and}\hspace{0.5cm} \delta=Re^{-\gamma}\to 0
\end{equation}
with the dimensionless temperature $\theta$ held constant.
Power $\gamma$ will usually remain on the range $3/4 \leq \gamma \leq 1$ as explained above.
On the other hand, power $\alpha > 0$ may vary depending on the limit protocol.
For instance, it is usual among mathematicians to consider the vanishing viscosity limit for the Cauchy problem with fixed initial data.
This means $\nu \to 0$ with all other parameters $U,L$, \textit{etc}, kept fixed.
This case would correspond to $\alpha = 1$.
Another approach~\cite{Eyink} would be to assume a non-vanishing mean energy dissipation rate per mass, satisfying Taylor's relation $\varepsilon\sim U^3/L$, which would lead to $\theta=\frac{2k_BT\varepsilon^{1/4}}{\rho\nu^{11/4}}$ and $\alpha = 15/4$.
This last approach is not far from the computed values for the ABL at $Re = 10^8$, which provides $\alpha = 4.875$ and $\gamma = 1.25$.
However, observe that the specific choice of power laws is not important as long as the limit is universal. We verify this universality numerically in this paper.

The phenomenon of spontaneous stochasticity, that may arise from the triple scaling limit~\eqref{triple_limit}, raises the following conjectures about the behavior of the system:
\begin{itemize}
    \item \textbf{Convergence:} The triple limit exists and is well-defined in a probabilistic sense.
    \item \textbf{Stochasticity:} The limit is a nontrivial stochastic process.
    \item \textbf{Spontaneity:} The resulting stochastic process is supported on the set of weak solutions of the Euler equations.
    \item \textbf{Universality:} The limit does not depend on the parameters  $\alpha$, $\theta$ and $\gamma$.
\end{itemize}

The goal of this paper is to address the above conjectures.
Direct numerical simulations of the full system~\eqref{NS_withNoise} are however computationally prohibitive.
We shall then rely on logarithmic lattices.

\section{Logarithmic lattice model}
\label{SEC:LogLatt}

\begin{figure}[t]
    \centering
    \includegraphics[width=5in]{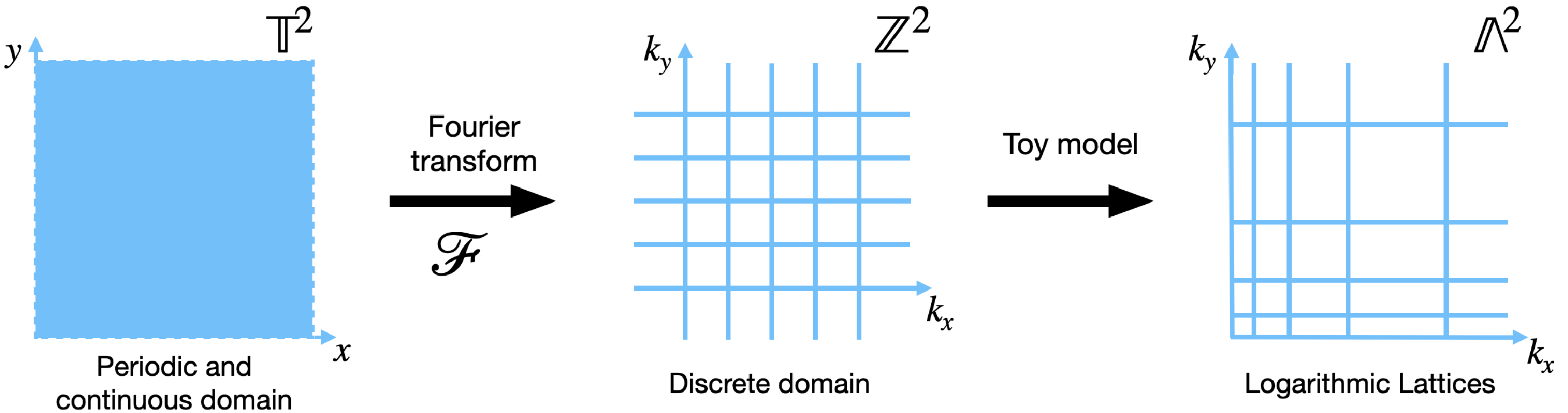}
    \caption{If we have a continuous and periodic domain (\textit{left}) and apply the Fourier transform, we obtain a discrete domain in the wave vector space (\textit{center}). Then, logarithmic lattices (\textit{right}) are toy models for the Fourier space, characterized by the fact that the spacing factor is no longer constant for all wave vectors. Instead, it increases geometrically as the wave vectors grow in norm.}
    \label{fig:LogLatt}
\end{figure}

A logarithmic lattice~\cite{LogL} is defined as the set
\begin{equation}
	\La=\{\pm \lambda^n\}_{n\in\Z},
\end{equation}
with a spacing factor  $\lambda>1$.
Each element $k\in\La$ is interpreted as a wavenumber in Fourier space, see Figure \ref{fig:LogLatt}.
The three-dimensional logarithmic lattice is given by the cartesian power $\La^3$, \textit{i.e.} $k = (k_1, k_2, k_3)\in \La^3$ if each component $k_j \in\La$.
Complex-valued functions $f(\vk)\in \C $ defined on a three-dimensional logarithmic lattice $\vk\in \La^3$ are considered analogous to the Fourier transform of a real-valued function.
For this reason, these functions satisfy the reality condition $f(-\vk)=f(\vk)^\ast$ and possess a natural structure of a real linear space.
Moreover, the spatial derivative $\partial_j$ in the $j$-th direction corresponds to the multiplication by the Fourier factor, \textit{i.e.} $\partial_jf(\vk)=ik_jf(\vk)$, where $i$ is the imaginary unit.
Higher-order derivatives are given by successive multiplications with such Fourier factors.

In this context, a new operation is introduced: the multiplication of two functions, which mimics the convolution in Fourier space of real-valued functions defined in Euclidean space. As a result, it reproduces many of the properties of the standard convolution---see \cite{LogL} for more details. The product is defined as follows,
\begin{equation}\label{prod}
    (f*g)(\vk)= \sum_{\substack{\bm{p} + \bm{q} = \vk\\[2pt] \bm{p},\bm{q} \in \La^3}}f(\bm{p})g(\bm{q}), \quad \vk \in \La^3.
\end{equation}
This operation defines local triadic interactions on the logarithmic lattice,  which are nontrivial only  for specific values of $\lambda$, such as $\lambda=2$, $\lambda\approx 1.618$ (the golden number) and $\lambda\approx 1.325$ (the plastic number).
In this paper, we always use the golden mean.

Logarithmic lattices were first introduced in~\cite{Blowup} and were further mathematically developed in~\cite{LogL,campolina2025logarithmic}.
As a technique to obtain simplified models, its main feature relies on changing the configuration space, and not the equations.
Therefore, the structure of the original system is retained, and most of its properties carry over.
Most importantly, the geometric growth of Fourier modes reduces dramatically the computational burden for the numerical simulations.
For these advantages, logarithmic lattices have already found several physical applications~\cite{barral2023asymptotic,costa2023reversible,pikeroen2023log}.

With the tools introduced above, we can now formulate the Navier-Stokes-Landau–Lifshitz model on the logarithmic lattice.
We consider the velocity field $\vu(\vk,t) \in \C^3$ and the scalar pressure $p(\vk,t) \in \C$ with the wave vector $\vk \in \La^3$ and time $t \in \R$.
These and all fields satisfy the reality condition.
The equations remain analogous to the continuous formulation and are given by
\begin{equation}\label{LLNS_LogLatt}
    \partial_t \vu + \vu  * \nabla \vu = -\nabla p + \frac{1}{Re}\Delta \vu + \nabla \cdot \bm{\tau}, \qquad \nabla\cdot \vu=0,
\end{equation}
with the fluctuating stresses $\bm{\tau} = (\tau_{ij}) \in \C^{3\times 3}$ modeled as Gaussian random tensors with covariance
\begin{equation}\label{covarianza_loglatt} 
	\langle \bm{\tau}_{ij}(\vk,t)\bm{\tau}^\ast_{kl}(\vk',t') \rangle = \frac{\theta}{Re^\alpha} \left(\delta_{ik}\delta_{jl}+\delta_{il}\delta_{jk}-\frac{2}{3}\delta_{ij}\delta_{kl} \right) \delta_{\vk \vk'}\delta(t-t').
\end{equation}
The above equations are taken within a high wavenumber cut off $|\vk| < \Lambda$, which scales with $Re$ as $\Lambda \sim Re^\gamma$.
This means that the nonlinear term is projected onto the cut-off wavenumbers, or, equivalently, we consider the truncated logarithmic lattice---see~\cite{LogL} for this formulation.
As a consequence of the logarithmic lattice structure, which mimics many of the properties of real valued functions on continuous physical space, the justification and deduction of the logarithmic lattice model is identical to the original system---see \ref{AppA} for details.
Particularly, Eqs.~\eqref{LLNS_LogLatt}-\eqref{covarianza_loglatt} are stochastically well-posed~(\textit{cf.} \cite{flandoli2008introduction}).

In the infinite Reynolds limit $Re\to \infty$, we formally obtain the Euler equations on logarithmic lattices
 \begin{equation}\label{Euler_LogLatt}
	\partial_t \vu + \vu \ast \nabla \vu = -\nabla p, \quad \nabla \cdot \vu = 0,
\end{equation}
which retain many of the original system’s symmetries and conservation laws~\cite{LogL}.
The symmetries are
\begin{itemize}
    \item \textit{(Time translations)} $t \mapsto t + s$, for any  $s\in\R$. 
    \item \textit{(Space translations)} $\vu \mapsto e^{i\bm{k}\cdot\bm{\xi}}\vu$, for any $\bm{\xi}\in \R^3$.
    \item \textit{(Isotropy and parity)} $\vk,\vu \mapsto R\vk,R\vu$, where $R$ is any element of the group of cube symmetries.
    \item \textit{(Scale invariance)} $t,\vk,\vu \mapsto \lambda^{h-n}t, \lambda^n\vk,\lambda^{-h}\vu$, for any $h\in \R$ and $n \in \Z$, where $\lambda$ is the lattice spacing.
    \item \textit{(Time reversibility)} $t,\vu \mapsto -t,-\vu$.
    \item \textit{(Galilean invariance---when $0 \in \La$)}  $\vu \mapsto e^{i\vk\cdot \bm{v}t}\vu+\widehat{\textbf{v}}(\vk)$, for any $\textbf{v}\in\R^3$, where $\widehat{\textbf{v}}(\vk)$ is the constant velocity field on the lattice defined as $\widehat{\textbf{v}}(0)=\textbf{v}$ and zero for $\vk\neq 0.$
\end{itemize}
Moreover, strong solutions of the Euler equations on logarithmic lattice preserve in time the same invariants as the continuous Euler equations, namely the kinetic energy $E(t)$ and the helicity $H(t)$
\begin{equation}
	E(t) = \frac{1}{2}\sum_{\vk\in \La^3}|\vu(\vk,t)|^2, \qquad H(t) = \sum_{\vk\in \La^3}\vu(\vk,t) \cdot \bm{\omega}(\vk,t)^\ast,
\end{equation}
where $\bm{\omega} = \nabla \times \vu$ are the vorticities.
Even circulation on loops (Kelvin's Theorem) is preserved in a generalized form~\cite{LogL}.

A necessary condition for spontaneous stochasticity is the low regularity and the non-uniqueness of solutions for the inviscid system.
For the Euler equations on logarithmic lattices~\eqref{Euler_LogLatt}, it was proved~\cite{LogL} local-in-time existence of a unique strong solution.
More precisely, if the initial condition $\vu^0$ has finite norm
\begin{equation}\label{hm_norm}
	\Vert \vu^0 \Vert_{h^m}^2 = \sum_{\vk\in \La^3} |\vk|^{2m} |\vu^0(\vk)|^2 < \infty, \qquad m \geq 1,
\end{equation}
then there exists a unique solution $\vu(t)$ continuous in time with values in $h^m$, \textit{i.e.} with finite norm~\eqref{hm_norm}, whose initial condition is $\vu \big|_{t=0} = \vu^0$.
Existence of such solution is guaranteed only for a finite interval of time.
A Beale-Kato-Majda blowup criterion~\cite{LogL} states that either the solution exists globally, or there is a finite time $0< t_b < \infty$ (called blowup time) for which
\begin{equation}
	\int_0^{t_b} \Vert \bm{\omega}(t) \Vert_{\infty} dt = \infty,
\end{equation}
where $\Vert \bm{\omega}(t) \Vert_{\infty} = \sup_{\vk \in \La^3} | \bm{\omega}(\vk,t) |$ is the supremum norm on the logarithmic lattice.
In~\cite{Blowup,LogL}, robust numerical evidence shows that a finite-time singularity occurs for rather general class of regular initial conditions.

In this context, if the initial conditions are regular enough, there should not be spontaneous stochasticity as long as the solution remains regular.
After the finite-time blowup, the solution loses regularity, and uniqueness is no long guaranteed, setting out a possible scenario for spontaneous stochasticity.
Such dichotomy was already successfully confirmed in the case of shell models of turbulence~\cite{SptS1}, where strong deterministic convergence was observed before blowup, and spontaneous stochasticity sets up after the singularity.

To solve numerically the equation~\eqref{LLNS_LogLatt}, we will consider its vorticity formulation
\begin{equation}\label{vorcityF}
\partial_t \vw  = \vw *\nabla \vu- \vu * \nabla\vw + \frac{1}{Re} \D \vw + \nabla\times(\nabla \cdot \bm{\tau}).
\end{equation}

This avoids the need to explicitly calculate the pressure field during time integration.
The velocity field can subsequently be recovered from the vorticity using the Biot-Savart law $\vu=\frac{i\vk \times \vw(\vk)}{|\vk|^2}$, while the pressure, if needed, can be obtained by solving the Poisson equation $-\Delta p=\nabla \cdot (\vu * \nabla \vu)$.
Our numerical strategy for solving the equation \eqref{vorcityF} is based on a split-step technique, considering three different steps.
First, we solve the nonlinear term using a classical 4th-order Runge-Kutta method~(\textit{cf.}~\cite[\S 5.7]{leveque2007finite}).
Next, we solve the noise term using Euler-Maruyama method~\cite{higham2001algorithmic}.
Finally, we solve the dissipative term using a low-pass filter (\textit{cf.}~\cite{LogL}).
The truncation cutoff and the step size are chosen to avoid numerical instability. These parameters depend on the Reynolds number and are chosen empirically by testing several simulations. In this process, we compare the spectrum of energy of these simulations and select the cutoff that allows proper energy propagation across scales, ensuring the energy cascade down to the smallest scales, while avoiding excessively small-scale spectrums with litte contribution to the dynamics and higher computational cost.
The total number of grid points depends on the chosen cutoff wavenumber. Specifically, for a cutoff wavenumber $\Lambda = \lambda^{N-1}$, the total number of grid points is $4N^3$.
All operations on logarithmic lattices are performed using LogLatt~\cite{campolina2020loglatt}.

\section{Results}
\label{SEC:results}

We study the Cauchy problem for~\eqref{LLNS_LogLatt} in two different setups.
In the first, we shall consider an irregular initial condition, \textit{i.e.}, with infinite norm~\eqref{hm_norm}.
In this case, there is no strong solution, and only weak (non-unique) solutions are expected.
Spontaneous stochasticity can arise from time $t = 0$.
In the second setup, we consider a regular initial condition, with finite norm~\eqref{hm_norm}.
Now, a strong solution exists for a short period of time.
We shall verify numerically the blowup at finite time $t_b$, and study the spontaneous stochasticity for instants $t > t_b$.

In both cases, we consider power-law initial conditions of the form
\begin{equation}\label{power_law}
	\vu^0(\vk) \sim |\vk|^{-\xi}, \quad \text{for} \ \xi > 0.
\end{equation}
The degree of regularity of the initial condition depends on the exponent $\xi$.
One may verify that if $\xi > 1$, then the initial condition is regular.
In this case, a strong solution exists locally in time.
On the other hand, if $\xi \leq 1$, then the initial condition is irregular.
Observe, however, that any power-law initial condition of form~\eqref{power_law} has finite energy.

To probe spontaneous stochasticity, we need to measure statistics.
In what follows, our main observable will be the probability density function of the vorticity field, evaluated at a fixed large scale wave vector and at a fixed instant of time.  We performed this statistical analysis for different large-scale wave vectors and obtained similar results. Therefore, in the following we specify the wave vector at large scales as $k_* = (\lambda,1,1)$ and consider the real part of the first vorticity component. 
We remark that if weak convergence is taking place, the PDFs should converge at all wave vectors.
Nevertheless, they have different rates of convergence.
We have then chosen wave vectors where the convergence was quicker when compared to others.
The same results are, however, expected for all wave vectors, possibly requiring higher Reynolds numbers.

\subsection{Irregular initial condition}

We consider the rough initial condition
\begin{equation}
    u^0_1(\vk) = u^0_2(\vk) =e^{i \phi(\vk)}|\vk|^{-1/2},
\end{equation}
where the phases $\phi$ are given by $\phi(\vk)= \sqrt{2}k_x  + \sqrt{3}k_y + \sqrt{5}k_z$, and the third component of velocity is uniquely defined by the incompressibility condition. We have chosen these phases such that they are reasonably "random" and reproducible.
\begin{figure}[!h]
    \centering
    \includegraphics[width=1\textwidth]{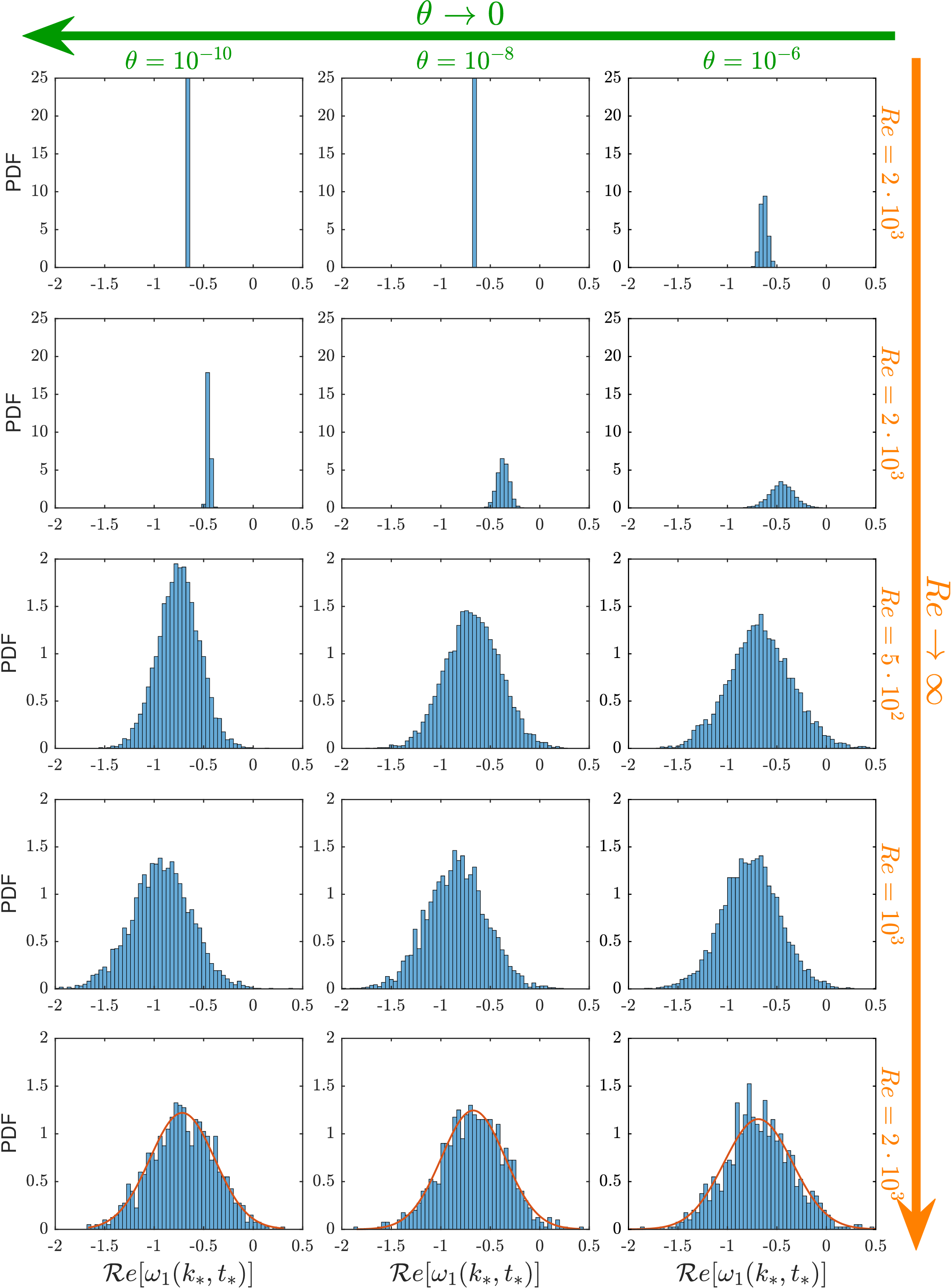}
    \caption{Probability density functions (PDFs) of the real part of the first component of the vorticity field, evaluated  at a fixed large scale wave vector  $k_* = (\lambda,1,1)$ and time $t_\ast=0.1$, for different values of Reynolds numbers and dimensionless temperature $\theta$. Solid lines in the last row represent the Gaussian PDFs with the same mean and variance.}
    \label{fig:ICI}
\end{figure}
The roughness of the initial condition allows us to avoid waiting for a blowup and instead promoting the rapid manifestation of multiscale effects and the loss of uniqueness of solutions.
In this way, we can detect and analyze spontaneous stochasticity with reduced computational requirements.
For this case, we set $\alpha = 1$, which corresponds to the abstract mathematical vanishing viscosity limit $\nu \to 0$ for the Cauchy problem with fixed initial data.

In Figure~\ref{fig:ICI}, we observe the PDFs for the real part of the first component of the vorticity field, evaluated at a fixed large-scale wave vector $k_* = (\lambda,1,1)$ and at time $t=0.1$.
The varying parameters are the Reynolds number $Re$  and the dimensionless temperature $ \theta$. In these cases, we consider large but finite values for the Reynolds number, specifically $10^2, \, 2\cdot 10^2, \, 5\cdot 10^2, \, 10^3 $ and $2\cdot 10^3$ and small but positive values of $\theta$, namely $10^{-6}, \, 10^{-8}$ and $10^{-10}$.  Additionally, we vary the cutoff wavenumber, setting it to $\lambda^{16} \approx 2 \times 10^3$ for the lowest Reynolds number, $\lambda^{24} \approx 10^5 $ for the highest Reynolds number, and $\lambda^{20} \approx 9 \times 10^3$ for the others.

If we fix the Reynolds number $Re$ and let the dimensionless temperature decrease, the PDFs collapse to a Dirac delta distribution in the limit $\theta \to 0$.
This behavior is expected: in this limit, the thermal noise vanishes, and the system reduces to the deterministic truncated Navier–Stokes equations at fixed $Re$, which admit a unique solution.
 However, since the noise amplitude also depends on  $Re$, the rate of this convergence is strongly influenced by the Reynolds numbers. This can be observed in Fig. ~\ref{fig:ICI}. For small values of $Re$, we can observe convergence to a Dirac delta distribution. In contrast, for larger Reynolds, this collapse is not visible within the range of $\theta$ values we tested. This does not mean that there is not convergence to a Dirac delta, instead, it indicates that the convergence is slower, and that even smaller values of $\theta$ would be required in order to observe it.

The more interesting case arises when we fix $\theta$ and increase $Re$.
We recall that the noise amplitude scales as $Re^{-\alpha/2}$, and for this case, $\alpha=1$, meaning that the stochastic forcing becomes smaller as $Re$ increases.
As we can see in Fig.~\ref{fig:ICI}, the PDFs converge to a non-delta distribution, indicating that the solutions remain stochastic even in the vanishing-noise and vanishing-viscosity limit.
The limiting distributions coincide up to statistical fluctuations for distinct values of $\theta$, suggesting universality. 
Solid lines in the last row of Fig.~\ref{fig:ICI} show the Gaussian PDFs with the same mean and variance,  suggesting that the large-scale distributions are close to normal.

Since the amplitude of the noise is small (reaching a minimum value of approximately $10^{-7}$) and concentrated at small scales, we conclude that the observed order $\mathcal{O}(1)$ stochastic behavior at large scales cannot be a direct consequence of the finite-size thermal forcing.
Instead, it reflects the nonlinear backward amplification of infinitesimally small noise from small to large scales.
This is the mechanism of spontaneous stochasticity. 

\subsection{Regular initial condition}

\begin{figure}[t]
	\centering
	\includegraphics[width=1\textwidth]{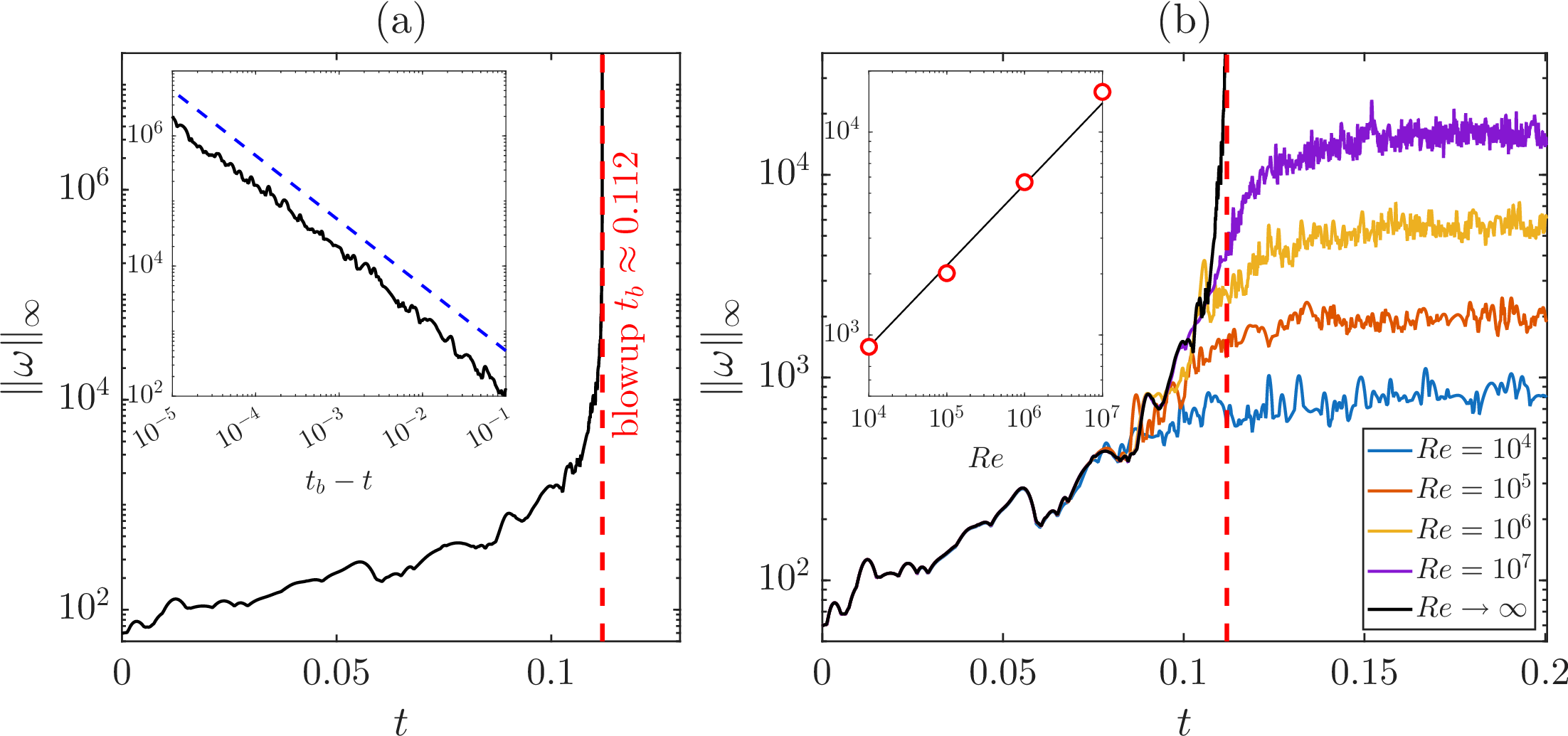}
	\caption{Time evolution of the supremum norm of the vorticity $\Vert \bm{\omega}(t)\Vert_{\infty}$. The vertical axis is plotted using logarithmic scale, which allows us to observe the growth of the norm across several orders of magnitude. In the left panel, we see the rapid growth of the vorticity norm for the solution of the deterministic Euler equations, exhibiting the asymptotics $\Vert \bm{\omega}(t) \Vert_\infty \sim (t_b-t)^{-1}$ as it approaches the blowup time $t_b \approx 0.112$. The solid line in the inset shows the same graph in loglog scale with the horizontal coordinate $t_b-t$, compared to the scaling $\sim (t_b-t)^{-1}$ shown by the dashed line. In the right panel, we can see the evolution of the same vorticity norm for the solutions of the deterministic Navier-Stokes equations at different Reynolds numbers.
    The inset shows final-time values as functions of $Re$ in loglog scale compared to the power law $\propto Re^{0.4}$.}
	\label{fig:Blowup}
\end{figure}

In order to improve convergence without requiring a large number of simulations and inspired by the symmetry groups of the Euler equations on the logarithmic lattice, we constructed an initial condition invariant under certain rotations. We observe that each of these rotations applied to the solution generates a different solution for the same initial data but with a different realization of noise. In this way, a single simulation produces multiple distinct solutions, improving statistical analysis with a reduced number of simulations. Additionally, to ensure a certain regularity, we impose that
\begin{equation}
	\vu^0(\vk) \sim |\vk|^{-3/2}.
\end{equation}
Thus, our initial condition will be the Leray projection of a rotation-invariant power law onto the solenoidal vector fields---see~Eq.~\eqref{Leray} in~\ref{AppA}.  
In this case, we take $\alpha = 15/4$, which is consistent with Taylor's relation of non-vanishing mean energy dissipation rate in the infinite Reynolds limit $Re \to \infty$~\cite{Eyink}.

Given the regularity of the initial condition, we expect a finite-time blowup in the deterministic Euler equations~\eqref{Euler_LogLatt}, as previously observed in~\cite{Blowup}.
To determine the blowup time, we solve the Euler equations and apply the Beale-Kato-Majda criterion.

The singular behavior is confirmed in Fig.~\ref{fig:Blowup}, which consists of two panels.
In the left panel, we observe the rapid growth of the vorticity norm for the solution of the Euler equations, exhibiting asymptotic behavior $\Vert \bm{\omega} \Vert_\infty \sim (t_b - t)^{-1}$, for the estimated blowup time $t_b=0.112$. 
In the right panel, we can see the evolution of the same vorticity norm $||\omega||_{\infty}$ but for the solutions of the Navier-Stokes equations at different Reynolds numbers, namely $10^4, \, 10^5, \, 10^6, \, 10^7 $.  And with the same cutoff wavenumber $\lambda^{39}\approx 1.4 \times 10^8.$
The Euler solution is plotted here as the limit $Re \to \infty$, for reference. And the cutoff wavenumber that we are considering is $\lambda^{59}\approx 2\times 10^{12}$.

\begin{figure}[t]
	\centering
	\includegraphics[width=1\textwidth]{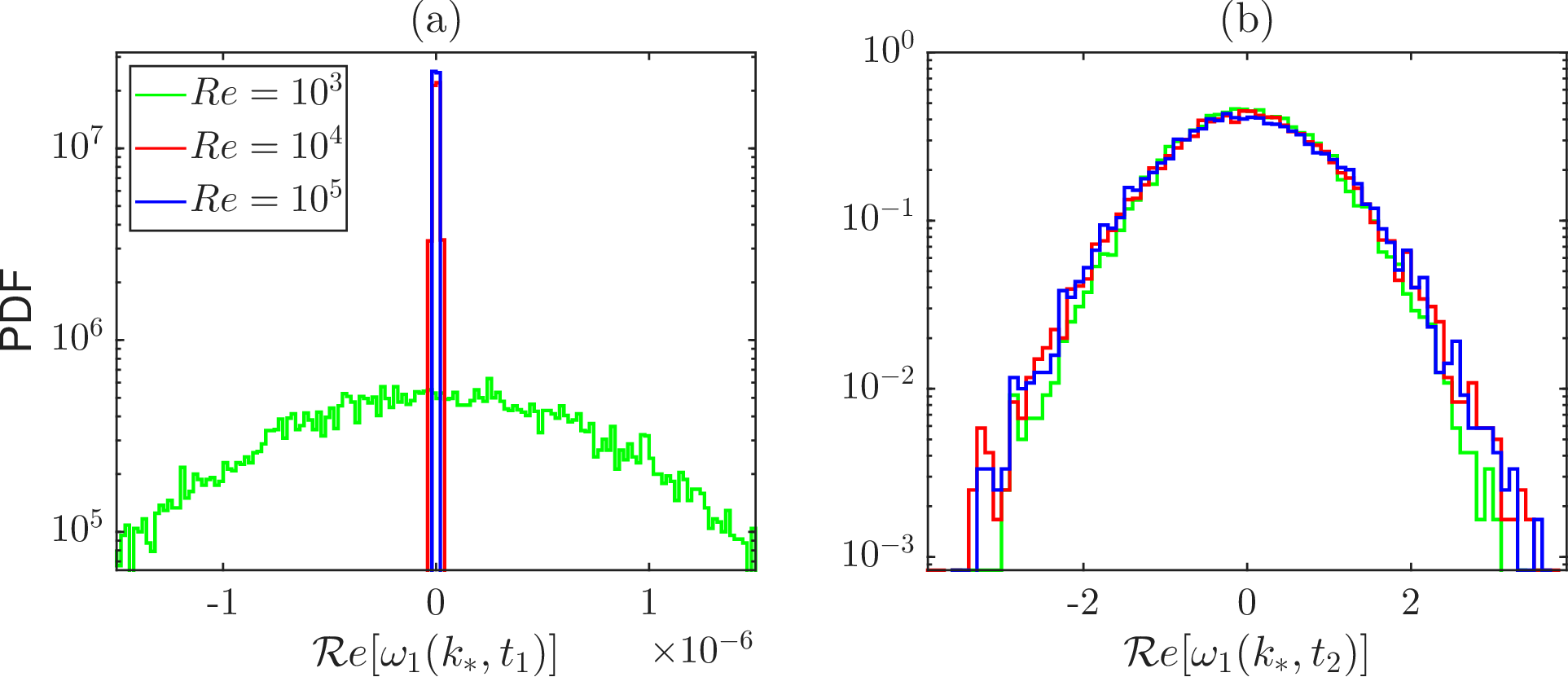}
	\caption{PDFs of the real part of the first component of the vorticity field, evaluated  at a fixed large scale wave vector  $k_* = (\lambda,1,1)$ and at two different instants of time, $t_1=0.075$ (before Euler's blowup) and $t_2=0.2$ (after Euler's blowup), in order to illustrate the two different statistical behaviors in the solutions of our model. The left panel corresponds to $t_1$, where we observe convergence to a deterministic solution. In contrast, the right panel correspond to $t_2$, where convergence to a spontaneously stochastic solution takes place.}
	\label{fig:ICR}
\end{figure}

For $t<t_b$, as the Reynolds number increases, the vorticity norm converges to the Euler solution.
Such convergence is expected from classical theory of strong solutions of Navier-Stokes and Euler equations~\cite{kato1972nonstationary}.
Observe that the convergence gets better and closer to the blowup time as we increase $Re$.

For any finite value of $Re$, the Navier-Stokes solutions can overpass the inviscid blowup time.
The vorticity norm saturates to an approximately constant mean value.
This is a consequence of the regularizing effect of viscosity at small scales.
The saturation value increases with $Re$ as $\Vert \bm{\omega} \Vert_\infty \sim Re^{0.4}$, a power-law dependence very close to the one found in~\cite{pikeroen2024tracking}.
This means that the vorticity blows up in the infinite Reynolds limit.

In Fig.~\ref{fig:ICR} we can see the PDFs of the real part of the first component of the vorticity field, evaluated at a fixed large-scale wave vector $k_* = (\lambda,1,1)$ and at two different instants of time, $t_1=0.075 < t_b$ (before Euler's blowup) and $t_2=0.2 > t_b$ (after Euler's blowup).
In this case, the dimensionless temperature has the fixed value $\theta=10^{-6}$, while the Reynolds numbers have values $10^3, \, 10^4$ and $10^5$. This implies that the smallest noise amplitude is given by $10^{-13}$. 
In the left panel, we have the PDFs before Euler's blowup.
We can observe that the PDFs converge with increasing $Re$ to a Dirac delta distribution.
This behavior is consistent with Fig.~\ref{fig:Blowup}, where convergence of the vorticity norms is observed for $t<t_b$ as the Reynolds number increases.
In the right panel, we have the PDFs after Euler's blowup.
Now we observe that the PDFs no longer approach a delta function.
Instead, for all three Reynolds numbers we have non‑trivial distributions with finite width, which converge as  $Re$ increases.

Therefore, we conclude that the solutions are converging towards a spontaneously stochastic solution. This stochasticity is not a result of the finite-size noise term into the model, but an intrinsic feature of the dynamics itself, which is able to amplify an infinitesimal small-scale noise to large values at large scales.
This provides numerical evidence that, after blowup, the system remains stochastic even in the vanishing-noise limit.

\section{Conclusions}
\label{SEC:conclusions}

In this work, we have studied the phenomenon of spontaneous stochasticity in the 3D incompressible Landau-Lifshitz-Navier-Stokes equations for fluctuating hydrodynamics on a logarithmic lattice.
To carry out this study, we have used probability density functions (PDFs) to analyze the statistics of the solutions at a fixed time and wave vector.
The Reynolds number was increased in order to emulate the limit $Re \to \infty$.
The noise covariance amplitude $\Theta$ and the wavenumber cut off $\Lambda$ were scaling with $Re$ as $\Theta = \theta Re^{-\alpha} \to 0$ and $\Lambda \propto Re^\gamma \to \infty$.
Focusing on large scale wave vectors, where we have faster weak convergence, we observed spontaneous stochasticity in two different setups.
In the first, a spontaneous stochastic solution arises immediately from a rough initial condition.
In the second, we consider regular initial data, for which a strong solution for the inviscid deterministic system exists up to a finite-time blowup.
Before the blowup time, the limiting solution is the deterministic strong Euler solution.
After the blowup time, the limit remains stochastic.
In both setups, the spontaneous stochastic solutions are characterized by finite positive variance of PDFs.
The fact that the statistics of solutions do not depend on $\theta$, and that the phenomena was verified for two rather different values of $\alpha$, points towards universality.
The considerable small values of $Re^{-1}$ and noise amplitude reached in our simulations show that we are approaching weak Euler solutions.
We have thus provided numerical evidence for the four raised conjectures about spontaneous stochasticity, namely: convergence, stochasticity, spontaneity and universality.

We expect that the formulated logarithmic lattice model might be a useful testing ground for future hypotheses and studies on spontaneous stochasticity in the Navier-Stokes system.

\vskip6pt

\ack{The authors thank B. Dubrulle for the useful discussions and suggestions. E.O. acknowledges the  financial support received during her visit to CEA-Université Paris-Saclay and ENS-Lyon, where her work was partially conducted. The research of C.C. was funded by the French National Research Agency (ANR Project TILT ANR-20-CE30-0035) and the European Union (ERC NoisyFluid, No. 101053472). A.A.M. was supported by CNPq grant 308721/2021-7, FAPERJ grant E-26/201.054/2022, and CAPES MATH-AmSud project CHA2MAN.}

\appendix
\renewcommand{\thesection}{Appendix \Alph{section}}

\section{Landau-Lifshitz-Navier-Stokes equations}\label{AppA}

We remind here a brief formal deduction of the thermal noise forces in the Landau-Lifshitz-Navier-Stokes equations.

Equation~\eqref{covarianza} arises from the motion of individual molecules in the medium due to their thermal energy.
These molecular interactions occur randomly and uniformly throughout the fluid, meaning the system does not exhibit any preferred spatial direction or orientation.
Consequently, the stochastic fluctuations in the stress tensor induced by thermal motion are modeled as isotropic and as both temporally and spatially uncorrelated.
We model it as Gaussian white noise, whose statistical properties remain invariant under spatial rotations and translations.
The most general form of a rank 4 isotropic tensor, which has no temporal or spatial correlations, is given by
\begin{equation}
	\langle \bm{\tau}_{ij}(\vx,t)\bm{\tau}_{kl}(\vx',t') \rangle = (a\delta_{ik}\delta_{jl}+b\delta_{il}\delta_{jk}+c\delta_{ij}\delta_{kl}) \delta^3(\vx-\vx')\delta(t-t').
\end{equation}
The coefficients $a, \, b$, and $c$ are determined by imposing additional physical constraints derived from fundamental conservation laws. First, since thermal fluctuations arise from random molecular motion that do not introduce any net torque, they must preserve angular momentum. This requirement implies that the noise tensor must be symmetric, which results in $a=b$.
Additionally, these fluctuations occur in an incompressible flow, meaning that they do not modify the local volume of fluid, and thus, the noise tensor is traceless.
Hence $c=-\frac{2}{d}a$, where $d=3$ is the spatial dimension.
In this way, the covariance of the stress tensor depends on a single parameter and takes the form
\begin{equation}
	\langle \bm{\tau}_{ij}(\vx,t)\bm{\tau}_{kl}(\vx',t') \rangle = a\left(\delta_{ik}\delta_{jl}+\delta_{il}\delta_{jk}-\frac{2}{3}\delta_{ij}\delta_{kl} \right) \delta^3(\vx-\vx')\delta(t-t').
\end{equation}

Following the approach in~\cite{Diss-Rate}, this remaining constant is determined using the Gibbs measure.
To achieve this, we first take the Fourier transform of equation (\ref{NS_withNoise}), restricted to a truncated Fourier space $K$ such that $|\vk|<\Lambda$ for all $\vk\in K$, and then apply the Leray projection operator in Fourier space, defined as
\begin{equation}\label{Leray}
	\proj_{ij}(\bm{k}) = \delta_{ij} - \frac{k_i k_j}{|\vk|^2},
\end{equation}
which projects onto the space of divergence-free vector fields in Fourier space. As a result, the fluctuating Navier-Stokes equation are reduced to a Langevin-type equation for the velocity modes 
\begin{equation}\label{Lang}
	\partial_t \Hat{\vu}_\vk = B_\vk[\Hat{\vu},\Hat{\vu}^*] -\frac{1}{\re} |\vk|^2 \Hat{\vu}_\vk +\Hat{\vq}_\vk,
\end{equation}
where $\Hat{\vu}_\vk$ represents the Fourier transform of the velocity field on the wave vector $\vk$ and $\hat{\vu}^*$ is its complex conjugate. The term $B_\vk[\Hat{\vu},\Hat{\vu}^*]=\proj [(\Hat{\vu}*\nabla)\Hat{\vu}]$ represents the Leray projection of the nonlinear term of the Navier-Stokes equation, while $\Hat{\vq}_\vk$ is the Fourier transform of the thermal noise.

From now on, we omit the hats on the variables to simplify the notation, with the understanding that all fields are expressed in Fourier space. 
Due to the reality condition, which implies $\vu(-\vk) = \vu^*(\vk)$, the dynamics can be restricted to a subset $K^+$ of independent wave vectors. This restriction ensures that only distinct wave vectors are considered, simplifying the representation of the system in Fourier space.
Later, introducing the notation $V_\vk[\vu,\vu^*] = B_\vk[\vu,\vu^*]-\frac{1}{\re}|\vk|^2\vu_\vk$ and  noting that the same fluctuation–dissipation covariance applies in Fourier space by replacing position $\vx$ with wave vector  $\vk$,  the Fokker–Planck equation associated with the system of coupled Langevin equations~\eqref{Lang} can then be written as
\begin{equation}
	\partial_t P =\sum_{\vk\in K^+}\Big[-\frac{\partial}{\partial \vu_\vk}\cdot (V_\vk[\vu,\vu^*]P)-\frac{\partial}{\partial \vu^*_\vk}\cdot (V^*_\vk[\vu,\vu^*]P) + a \proj : \frac{\partial^2}{\partial \vu_\vk \partial \vu^*_\vk}\Big].
\end{equation}

Now, to determine the constant $a$, we replace the probability density function 
$P$ with the Gibbs measure in dimensionless form
\begin{equation}
	P_G[\vu,\vu^*]=\frac{1}{Z}\exp{\bigg(-\frac{\rho U^2 L^3}{k_B T V}\sum_{\vk\in K^+}|\vu_\vk|^2\bigg)}.
\end{equation}
Since the fluctuating hydrodynamics equation preserves the Gibbs invariant measure, then $\partial_t P_G =0$, and, using the Liouville theorem on conservation of phase-volume and the conservation of kinetic energy, it follows that $a$ has the value of $\Theta$ we have considered in Eq.~\eqref{Re_Theta}.
For more detailed derivations and explanations, see the Appendix A of \cite{Diss-Rate}.

We remark that the above derivation is entirely conducted in Fourier space, and it also holds for the logarithmic lattice model.

\bibliographystyle{RS} % estilo de bibliografía: puedes usar alpha, ieee, apalike, etc.
\bibliography{sample} % nombre del archivo .bib (sin extensión)

\end{document}